\documentclass[letterpaper,12pt]{article}

\usepackage{graphics,graphicx}
\usepackage{times}
\usepackage{wrapfig}
\usepackage{natbib}

\usepackage{amsmath}
\usepackage{comment}
\usepackage{xcolor}
\usepackage{xcolor}
\definecolor{amethyst}{rgb}{0.6, 0.4, 0.8}

\newcommand{\solphys}{Solar~Phys.}
\newcommand{\apj}{Astrophys.~J.}
\newcommand{\apjs}{Astrophys.~J.~Suppl.}
\newcommand{\apjl}{Astrophys.~J.~Letters}
\newcommand{\mnras}{Monthly Notices Royal Astronom. Soc.}
\newcommand{\araa}{Ann.~Review~Astron.~Astrophys.}
\newcommand{\aap}{Astron.~Atrophys.}
\newcommand{\ssr}{Space Sci.~Rev.}

\newcommand{\procspie}{Proc. SPIE}
\newcommand{\physrep}{Physics Reports}
\makeatletter
\renewcommand{\section}{\@startsection%
{section}{1}{0mm}{-\baselineskip}%
{0.5\baselineskip}{\normalfont\Large\bfseries}}%
\makeatother

\begin{document}
\pagestyle{plain}
\pagenumbering{arabic}

\centerline{\bf MULTI-HEIGHT MEASUREMENTS OF THE SOLAR VECTOR MAGNETIC FIELD}

\bigskip

A White Paper Submitted to the Decadal Survey for Solar and Space Physics (Heliophysics)
2024 - 2033

\vspace{2cm}

{\bf L. Bertello}$^1$, 
N. Arge$^2$, 
A. G. De Wijn$^3$,
S. Gosain$^1$,
C. Henney$^4$,
K.D. Leka$^5$,
J. Linker$^6$,
Y. Liu$^7$,
J. Luhmann$^8$,
P.J. Macniece$^2$,
G. Petrie$^1$,
A. Pevtsov$^1$,
A.A. Pevtsov$^1$

\bigskip

$^1$National Solar Observatory, 
$^2$NASA/GSFC,
$^3$HAO/NCAR,
$^4$AFRL/Space Force, 
$^5$NWRA \& Nagoya University,
$^6$Predictive Science Inc.,
$^7$Stanford University,
$^8$UC Berkeley

\vspace{3cm}

{\bf Synopsis:}

\medskip

This white paper advocates the importance of multi-height measurements
of the vector magnetic field in the solar atmosphere. As briefly described in
this document, these measurements are critical for addressing  some of the most fundamental
questions in solar and heliospheric physics today, including: (1) What is the
origin of the magnetic field observed in the solar atmosphere? (2)
What is the coupling between magnetic fields and flows throughout the solar atmosphere?
Accurate measurements of the photospheric and chromospheric three-dimensional magnetic 
fields are required for a precise determination of the emergence and evolution of active regions. 
Newly emerging magnetic flux in pre-existing magnetic regions causes an increase in the 
topological complexity of the magnetic field, which leads to flares and coronal mass ejections.
Measurements of the vector magnetic field
constitute also the primary product for space weather operations, research, and modeling of
the solar atmosphere and heliosphere. \\
The proposed next generation Ground-based solar Observing Network Group (ngGONG), a coordinated
system of multi-platform instruments, will
address these questions and provide large datasets for statistical
investigations of solar feature behavior and evolution and
continuity in monitoring for space-weather focused endeavors both research and operational.
It will also enable sun-as-a-star investigations, crucial as we look toward understanding other
planet-hosting stars.

A wide-spread use of full-disk vector magnetic field observations in solar physics research, 
their increasing importance for understanding many fundamental phenomena, and a 
growing potential of these data for operational space weather forecast strongly suggest that 
these type of observations need to be continued as part of long-term (synoptic) 
program from ground-based and/or space-based facilities. Regular multi-height observations of 
magnetic fields on the Sun is the next frontier in Solar Physics.

\newpage

\section{Introduction}

The Sun dominates the Earth's climate and the space environment
throughout the solar system. It provides variable radiative, particle,  and
magnetic field input to the conditions in the heliosphere which
directly influence the Earth's magnetosphere. As a dominant source of
electromagnetic energy and energetic particles, we need to
understand the Sun's sources of variability. 

Observations of the solar surface reveal magnetic and velocity fields
with complex hierarchical structures, evolving on a wide range of different spatial and temporal
scales. These features are the direct manifestation of 
a hydromagnetic dynamo process operating in the Sun's interior,
and leading to the 11 (or 22) -year solar activity cycle 
\citep[e.g.][]{2014ARA&A..52..251C}.
The interaction between flow velocity, the motion of flux tubes through
the convection zone, and the generation of magnetic fields 
are issues which remain open and which are
critical to the successful understanding and modeling of the properties of the
solar/stellar dynamo that affect the solar atmosphere and active region
generation, and the related solar irradiance and heliospheric outflows that ultimately
result. Accurate measurements of the photospheric and chromospheric 
three-dimensional magnetic
fields are required for a more precise determination of the emergence and evolution 
of active regions,
wherein newly emerging magnetic flux increases the topological complexity, leading to 
flares and
coronal mass ejections. The challenge is to understand them well enough to be able 
to predict their
occurrence and characteristics.

Due to the availability of full disk vector magnetic field measurements, 
great progress has been made over the last decade to understand solar variability at 
different temporal and spatial scales. 
Because different spectral lines and/or different points
along the profile of a single spectral line are formed at different heights in the solar atmosphere,
multi-wavelength and spatially-resolved observations of the magnetic field constitute the only
tool available today for resolving the 3D structure of the magnetic field in the solar atmosphere.
Significant contributions to this effort have been  provided 
by the continuous observations by the Vector 
Spectro-Magnetograph \citep[VSM/SOLIS,][]{Keller.etal2003}, the Hinode/SP
\citep[e.g.][]{2013SoPh..283..579L}, and later 
by the Helioseismic and Magnetic Imager (HMI) instrument on the Solar Dynamics Observatory \citep{hoeksema2014,scherrer2012}. The high temporal cadence and continuity of the full-disk HMI vector magnetic field data available since 2010 provides a unique opportunity for a plethora of
 different studies about our star 
 \citep[e.g.][]{kazachenko2022,lumme2022}, and can be integrated with new technologies such as machine learning \citep[e.g.][]{higgins2022}. 
 Observations with the HMI instrument are taken at a single layer in the solar photosphere, but studies have shown that
 measurements of the magnetic field at multiple levels 
improves data interpretations such as transverse field disambiguation \citep{2009SoPh..260...83L}. 
They also minimize assumptions that have been used in various models such as non-linear force field model \citep{wiegelmann2008} and 
integral Lorentz force estimations
\citep{petrie2019}, that
 provide a better understanding of the morphology and dynamics in the region between the photosphere and chromosphere where both flow and field are equally important \citep{metcalf1995}.
This need also addresses
fundamental questions concerning the origin of the magnetic field and helicity observed in the
photosphere (deep seated vs. near surface dynamo activity or both), 
how magnetic helicity is stored in
the solar atmosphere and affects the magnetic energy available for solar eruptions, and the coupling
between magnetic fields and flows throughout the solar atmosphere, including the sites 
where the fast and slow solar wind are generated.

The reconciliation of the observed evolution of large-scale magnetic helicity and one predicted by
current solar dynamo models would require continuous observations of vector
magnetic field over the period of solar sunspot or even full magnetic cycle (20+ years).
Over the last decade significant advances in computational power, combined with the development 
of increasingly sophisticated numerical models, have made the full-disk measurements of the 
solar vector magnetic field especially relevant. When combined into synoptic charts or maps,
these measurements are used 
to drive the most advanced 3-D global numerical simulations 
\citep[e.g.,][]{2022SoPh..297..100B,2021JGRA..12628870L,2016usc..confE.101H}. 
The accurate construction and calibration of these maps establishes both
the diagnostic capabilities of the models and their ability to
forecast the state of the corona and heliosphere
\citep{2022ApJ...926..113W,2022SoPh..297...82F,
2007ApJ...667L..97R,2006SoSyR..40..432V,1995AdSpR..16i.181Z}. As examples, the
3-D geometry of the magnetic field expansion
in the inner corona, from the photosphere out to a few solar radii, plays
a fundamental role in determining  the density distribution and solar wind speeds
in heliospheric models, as the field lines define the flow tubes along which mass and energy
flux are conserved. This geometry is directly linked to the global topology of the solar
magnetic field, including that on the invisible far-side, and
in the polar regions where observations are limited.
Multi-height vector magnetic field measurements also play a 
critical role in addressing this particular issue of the farside and polar observational gaps
\citep{2021cosp...43E1793A,2021ApJ...917...27J,
2022ApJ...930...60H}.

For almost 20 years the Integrated Synoptic Program at the National Solar Observatory 
has provided to the solar community, through its SOLIS/VSM instrument, unique 
full-Stokes measurements of the full-disk photospheric magnetic field and more 
recently full- Stokes chromospheric measurements as well. 
This achievement is expected to be greatly enhanced by a newly proposed ngGONG 
(next generation GONG) project, a ground-based global network of highly capable solar 
observations to study the Sun and its consequences on Earth. 
Once operational, ngGONG will provide measurements of the processes that drive the 
activity from the solar interior, the atmosphere, and throughout the heliosphere.

The state-of-the-art  magnetohydrodynamic (MHD) coronal models require
full disk vector magnetic field information. 
In the near future, routine well-calibrated and sensitive
vector magnetic field measurements in both the photosphere and 
the chromosphere will be regularly required for modeling space weather, for both research and in 
support of operational forecasting. Improved performance 
will enable effective application of
the force-free condition adopted by some widely used models for 
extrapolating magnetic fields into the corona, and 
addressed the need for a robust approach to 
the 180-degree ambiguity resolution for the direction of the transverse field
\citep[e.g.,][]{2022SoPh..297...12V,2009SoPh..260...83L}. 
ngGONG will be designed to provide these required vector magnetic field data.

Multi-height measurements of the solar vector magnetic field also have important implications
for the using of helioseismology as a tool. 
For example, a study by \cite{2006MNRAS.372..551S} suggests that
the direction of the magnetic field plays an important role in how acoustic waves
can be converted into different types of magnetohydrodynamics (MHD) modes. That is, 
the magnetic field acts as a filter, preferentially allowing through acoustic signal from a 
narrow range of incident directions.  Simultaneous helioseismic and
multi-height magnetic observations will improve our
understanding of how acoustic wave propagate in the Sun, and help to clarify the complex
plasma structure and dynamics underneath active regions. This has important implications
in terms of forecasting the emergence of active regions and subsequent events associated
with space weather phenomena (e.g. flares, coronal mass ejection, etc...)

An important area in multi-wavelength studies of the Sun is the development of
interpretation tools for such data. Numerical inversion tools that can simultaneously fit 
multiple spectral lines (forming at different heights) and obtain a physically consistent 
3D magnetic and thermodynamic parameters are needed. 
Machine-learning tools in conjunction with realistic MHD and radiative transfer models 
can then be used to infer physical parameters in near real-time forecast applications.

However, there are research challenges to be met.  
Specifically, inversions that return the height-resolved atmosphere, or 
multi-line inversions that constrain the results even further are still not widely used, 
although they are being developed actively \citep{2016LRSP...13....4D}.  
In particular, for multi-height disambiguation of the 180$^{\circ}$ degeneracy and the further 
interpretation for physics-based inquiry (such as computing the spatially-resolveed 
Lorentz forces), the mapping from optical depth to physical height is required, 
and is not straightforward \citep{PastorYabar_etal_2019,Loptien_etal_2020}.

\section{ngGONG}

The next generation Ground-based solar Observing Network Group (ngGONG), a coordinated
World-wide system of multi-platform instruments, will provide a broad range of measurements
addressing the needs and goals described above for fundamental research in solar and space
physics, and space weather.
ngGONG will obtain observations of 
Doppler velocities in the photosphere, vector field magnetic fields in the 
photosphere and chromosphere, and full-disk spectrally-resolved scans of polarized light for 
Doppler velocity. 
The design of ngGONG is driven by the research interests of the national and international solar 
and solar-stellar scientific communities, and informed by the requirements of space-weather 
forecasting agencies. Led by the National Solar Observatory (NSO) in partnership with 
the National Center for Atmospheric Research (NCAR) High Altitude Observatory (HAO)
ngGONG will collect critical observations of solar activity and the Sun’s magnetic field 
over two Hale solar magnetic cycles (about 44 years). 
ngGONG will provide the information needed to
answer some fundamental questions about the nature of the solar
magnetic field, including:

\begin{itemize}
\item What is the origin of the magnetic field and helicity observed in the 
photosphere (deep seated vs. near surface dynamo activity or both)?
\item How is magnetic helicity stored in the solar atmosphere and how does it affect 
the amount of available magnetic energy and solar eruptions?
\item What is the coupling between magnetic fields and flows throughout the solar atmosphere?
\item Where are the the fast and slower solar wind generated?
\end{itemize}

More details about the design and
scientific objectives of ngGONG are provided in the White Paper 
"ngGONG -- Future Ground-based Facilities for Research in Heliophysics and Space Weather 
Operational Forecast" led by Alexei Pevtsov.

\section{A Science Case: Solar Magnetic Helicity}
\begin{wrapfigure}{r}{0.5\textwidth}
  \begin{center}
    \includegraphics[width=0.48\textwidth]{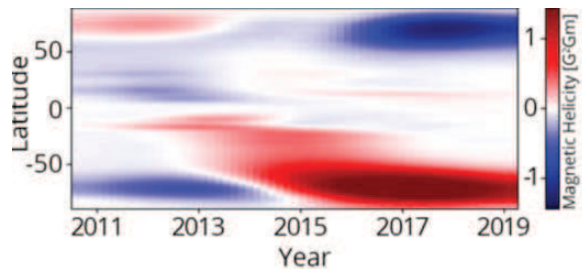}
  \end{center}
  \caption{ Puzzling sign-reversal of large-scale magnetic helicity in cycle 24  derived using vector magnetic field observations from SDO/HMI. During same period, small-scale helicity (not shown) exhibits ``regular'' hemispheric asymmetry without sign reversal as in the large-scale helicity. Adopted from \cite{Pipin.etal2019}.}
\label{heli}
\end{wrapfigure}

Magnetic helicity 
is an integral measure of topological properties of the magnetic field \citep[e.g.,][]{Pevtsov.etal2014}.
Locally, it can be characterized by a number of parameters such
as linkage, twist and writhe of the field lines \citep{2018JPhA51W5501B}.
In astrophysical dynamos, magnetic helicity is commonly accounted as a nonlinear constraint
of turbulent generation of large-scale magnetic field \citep{brsu05}.
Early studies found 
that the sign of magnetic helicity in solar
active regions follow the hemispheric helicity rule: predominantly negative values in the northern hemisphere and mainly
positive values in the southern hemisphere \citep[for review, see][]{Pevtsov.etal2014}.
However, recent studies indicate that the evolution of the magnetic helicity density of a
large-scale axisymmetric magnetic field is different from what is predicted by dynamo theory (Figure \ref{heli}). On one hand, the
mean large- and small-scale components of magnetic helicity density display the hemispheric helicity rule of
opposite signs at the beginning of cycle 24. However, later in the cycle, the two helicities exhibit the same sign, in
contrast with theoretical expectations \citep{Pipin.etal2019}.

Due to the large magnetic Reynolds number, the magnetic helicity originating from the solar 
interior can be carried away through the photosphere into the corona. In addition,
the kinetic helicity of subphotospheric flows is also 
a critical ingredient of the solar dynamo as it plays a role
in future flare activity of active regions after they crossed the visible solar surface. 
A deep-seated dynamo, which generates strong magnetic fields below the photosphere, 
is likely to be helical, while a near surface dynamo may not contain a net-helicity on a 
large scale. Comparing the kinetic helicity of subphotospheric flows with the 
magnetic field helicity provides invaluable information about flow of helicity through 
the solar atmosphere, and its eventual removal from the Sun - a key process for 
understanding the internal workings of solar dynamo. Currently there are two methods to estimate helicity in active regions. One is to integrate over time the helicity flux that ejects into the corona \citep[e.g.][]{liu2012}; the other is to calculate the helicity using modeled magnetic field in the active regions \citep[e.g.][]{valori2016}. The limitation for the former is that for many active regions, observation may not be available to catch the entire process of emergence and development so that the accumulated helicity calculated in this way may only contain a fraction of the total helicity in the regions; the latter depends on the modeled field in the volume of the active regions that might not precisely represent the real field.
More recently, \cite{2020ApJS..248....2F} and \cite{2020ApJS..250...28H}
have shown that full-disk vector magnetic field observations can
be used to derive the changing electric field in the solar photosphere over active-region scales.
Once the electric fields have been computed, it can be
use them to estimate the Poynting flux of energy in the radial direction
and the helicity injection rate contribution function.
However, because we do not have a sufficient amount of information, particularly on the 
vertical gradients of the
magnetic field around the photosphere, assumptions have to be made \citep{2018ApJ...855...11H}.

However, the 
relationship between the accumulated magnetic helicity flux through the photosphere and 
the magnetic helicity in the corona is still unclear.
Computation of magnetic helicity on the Sun requires knowledge of the vector magnetic field 
in a 3-D region. 
Measurements of the vector magnetic field at different heights
in the solar atmosphere are critical for a proper 
derivation of magnetic helicity on the Sun. 
Current observations are usually taken in a shallow layer of 
the solar atmosphere, typically in the photosphere, and a force-free parameter $\alpha$ is
commonly used as a proxy of magnetic helicity. To address this important limitation, ngGONG
will provide measurements of the magnetic vector field both in the photosphere and chromosphere.

The reconciliation of the observed 
evolution of large-scale magnetic helicity and one predicted by current solar dynamo models would require continuous observations of vector magnetic field over the period of solar sunspot or even full magnetic cycle (20+ years).


\section{Modeling: Space Weather Research and Operations}

It is clear that Space Weather research and operations will increasingly depend on 
the accuracy and availability of the
synoptic maps derived from magnetic field measurements that
are the main drivers of coronal and heliospheric models. 

There many factors that can affect the quality and diagnostic value of these maps. These
include high-noise level in the measured magnetic field, calibration
of the magnetic field, low-Zeeman polarimetric sensitivity, and difficulty in
in representing the true orientation of the solar magnetic fields with weaker polarization signals. 
For the specific case of full-Stokes measurements, the adopted inversion technique and
disambiguation method \citep{2021JSWSC..11...14P,2022SoPh..297...12V,Leka+2022}
will also affect the derived vector magnetic field.

However, one of the most critical limitations is represented by how poorly the polar regions
of the Sun are observed.
Polar field measurements are extremely important for several reasons: 1) they dominate the coronal
structure over much of the solar cycle (except when the polar fields reverse), 2) polar magnetic
flux plays a role in determining the properties/evolution of the heliospheric magnetic field, 3)
the polar magnetic fields are thought to be the direct manifestation of the Sun’s interior global
poloidal fields which serve as seed fields for the global dynamo that produces the toroidal fields
responsible for active regions and sunspots, and 4) the polar regions,
and their midlatitude extensions,  are the source of some of the fastest
solar wind streams \citep{2015LRSP...12....5P}.

Unfortunately, measuring the polar field is difficult due to foreshortening effects at 
the solar limb as
well as the intrinsic weakness of the field near the poles, and interpretation of these measurements
is complicated by a number of factors including the complexity of the polar magnetic landscape.
Hinode observations of the polar regions have revealed patches of magnetic field with different
spatial extent and distribution. While some are isolated, others form patterns like chains 
of islands.
Many of these patches are coherently unipolar and have field strengths reaching 
above 1 kG \citep{2008ApJ...688.1374T}. 
Their size tends to increase with latitude, up to about 5$\times$5 arcseconds. 
All of the large patches have fields
that are predominantly vertical relative to the local surface, while those of the smaller patches
tend to be horizontal. If a typical
radial correction is applied to line-of-sight  magnetograms, then
the horizontal fields are incorrectly amplified with a strongly varying radial function. 

Depending
on the distribution of the horizontal fields this may lead to a sign bias and inaccurate flux on any
given day.
Furthermore, for a given latitude, these effects will change with the B0 angle. Because of
projection effects, polar measurements obtained at favorable B0 angle (around March/September
for the southern/northern solar hemisphere) will be less noisy than other periods of the year.
The sensitivity of the magnetic field measurement is also a significant factor, and seeing plays a
role in ground-based observations. The impact of all of these factors
is expected to be greater during solar minimum, when the strength of the poloidal
field is stronger. 

There are different ways to improve the quality of synoptic magnetic maps used in
coronal and heliospheric models. From an observational point of view, better polarimetric
sensitivity (better signal-to-noise ratio) in measuring the transverse component of the solar 
vector magnetic field (Stokes Q and U) will improve the calculation
of the three components of the vector field. This is particularly important
for identifying the morphology of the magnetic field in regions away from the solar disk
center, where the transverse field is the dominant component of quiet/weak field regions. 
This includes the very important highest latitude regions located above 
approximately $\pm$60 degrees
latitudes. One of the major objectives of ngGONG is to produce such measurements.

\begin{wrapfigure}{l}{0.7\textwidth}
  \begin{center}
    \includegraphics[width=0.68\textwidth]{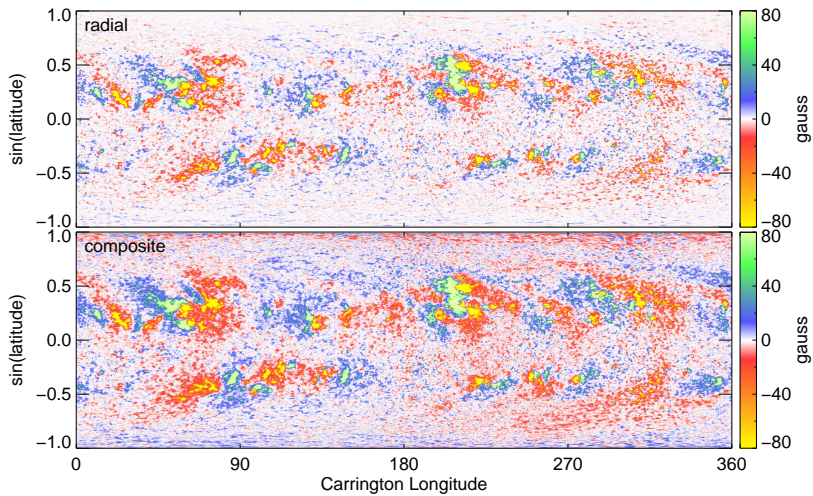}
  \end{center}
  \caption{Comparison between radial (top) and composite synoptic charts (bottom) 
derived from SOLIS/VSM FeI $\lambda630.2$ nm vector and longitudinal magnetic field observations
for Carrington rotation 2119 (Jan-Feb 2012).
Note the enhanced weak magnetic field in the composite chart.
}
\label{comp}
\end{wrapfigure}

Solar Orbiter, an international cooperative mission between ESA (the European Space Agency) 
and NASA, will observe the Sun's atmosphere up close, with high spatial resolution telescopes.
Thanks to its unique orbit, Solar Orbiter will also view the poles from solar latitudes higher than 
30 degrees (compared with $\sim$7 degrees from the Earth) and will 
provide an unprecedented picture of the magnetic environment in the Sun's polar regions.
These data will provide a unique opportunity to address one of the major
limitations in building current synoptic maps of the solar magnetic field. Since these
maps need to be fully filled with data, whenever one of the polar regions are not observable
from Earth some sort of extrapolation of the field at lower latitudes is adopted in order
to account for the missing data. Different techniques have been proposed, but there is
no consensus on which one is more reliable. 
Using these new data to first determine an optimum extrapolation from statistics of real
polar measurements, and then later routinely apply it to ngGONG maps, will reduce this
ambiguity.

Merging observations taken from different instruments, however, is not a trivial task. It
requires a carefully analysis of the properties of the individual data sets and a proper
cross-calibration scheme. A similar approach has been developed  at NSO to merge together
photospheric vector and longitudinal measurements taken by the SOLIS/VSM instrument. The
main purpose was to exploit the better sensitivity to the weak magnetic field provided
by the longitudinal measurement with the more reliable determination of the vector field
derived from the full-Stokes measurements. As illustrated in the bottom panel of Figure \ref{comp},
the result is a hybrid synoptic maps of the radial component of the magnetic field showing
an enhanced morphology of the field outside active regions. 
Further research is needed, however, to account for varying formation heights, 
the potential for multi-vantage-point and multi-line data, and 
determining the most appropriate representation for full 4$\pi$ boundary condition 
inputs for global models.

Growing sophistication in our knowledge, available (and possible) observations, and 
computational tools
have together fundamentally changed the way heliophysics is done, and how it fits into 
the broader
scope of NASA’s and other agency programs. In particular, both researchers and others interested
and/or invested in areas involving our space environment have become increasingly aware of, and in
many cases, users of, models. While empirical models have historically been a mainstay of 
applications
such as space weather event forecasting, there are new generations of these inspired by improved
understanding of underlying physics and processes. For example, modern flare and CME forecast models
now often rely on combinations of magnetograms and sometimes images to assess the likelihood of a
major x-ray outburst and/or a material eruption 
\citep[e.g.][]{2011SpWea...9.4003F,2020FrPhy..1534601G,LekaBarnesWagner2018,Leka_etal_2019}. These
approaches now can routinely make use of HMI images of active region vector fields combined with
chromospheric magnetic field observations and related modeling to reconstruct the field 
geometry \citep[e.g.][]{2020ApJ...896..119K}.
At the same time there is a long-standing modeling community goal toward
driving global models of the corona that include these active regions in their large scale 
context \citep[e.g. the review by][]{2017SSRv..210..249W}. 
Such a model would break through a critical barrier to
understanding how the solar activity cycle changes the character of the 
heliosphere 22 years, how it is
connected to the interior dynamo processes, and how coronal eruptive processes fit into the overall
dynamics of the entire coupled interior-to-heliopause system. 
The modeling work is moreover increasingly shared through outlets such as the
CCMC at GSFC, where a PFSS coronal field model is readily available, together with several 
coronal and
heliospheric MHD models, all currently based on synoptic photospheric magnetograms-but capable in
some of the latter cases of ingesting more detailed active region field data even today 
\citep[e.g.][]{refId0}.
It is important to recognize that all of these increasingly essential modeling efforts rely
on the availability and completeness of solar magnetic field observations, 
without which heliophysics would not be progressing.

\newpage


\end{document}